\documentclass[12pt,preprint]{aastex}

\newcommand{\vsig}{$ <v_{rot}^{obs}(Z)>/\sigma_{v}(Z)$}

\newcommand{\vsigmr}{$<v_{rot}>/\sigma_{v} $-metallicity relation}

\begin{document}
\title{Chemo-dynamical evolution of Globular Cluster Systems}
\author{Yoshihiko Saito and  Masanori Iye\altaffilmark{1} }
\affil{National Astronomical Observatory Japan, 2-21-2 Osawa, Mitaka, Tokyo 181-8588 Japan}
\altaffiltext{1}{Graduate University for Advanced Studies, Mitaka, Tokyo 181-8588 Japan}
 
\begin{abstract}

We studied the relation between the ratio of rotational velocity to velocity dispersion and the metallicity (\vsigmr ) of globular cluster systems (GCS) of disk galaxies by comparing the relation predicted from simple chemo-dynamical models for the formation and evolution of disk galaxies with the observed kinematical and chemical properties of their GCSs.

We conclude that proto disk galaxies underwent a slow initial collapse that was followed by a rapid contraction and derive that the ratio of the initial collapse time scale to the active star formation time scale is $\sim 6$ for our Galaxy and $\sim 15$ for M31. 

The fundamental formation process of disk galaxies
was simulated based on simple chemo-dynamical models assuming the conservation of their angular momentum. 
We suggest that there is a typical universal pattern in the \vsigmr \ of the GCS of disk galaxies. This picture is supported by the observed properties of GCSs in the Galaxy and in M31.
This relation would deviate from the universal pattern, however, if large-scale merging events took major role in chemo-dynamical evolution of galaxies and will reflect the epoch of such merging events.
We discuss the properties of the GCS of M81 and suggest the presence of past major merging event.

\end{abstract}


\section{Introduction}

The importance of investigations of globular clusters (GCs) has been reassessd in recent years (Ashman \& Zepf 1998, Kissler-Patig 2000),     
particularly since the globular cluster systems (GCS) of disk galaxies can be used to study the dynamical formation history of their host galaxies.

The rapid dynamic formation scenario of our Galaxy was first proposed by Eggen, Lynden-Bell \& Sandage (1962; hereafter ELS) based on apparent correlations among the kinematic properties and metallicity indicators of the halo and disk stars in the solar neighborhood.

Previous studies interpreted the properties of the GCS in relation to the formation of our Galaxy.
Maline, Hartmann \& Mathews (1991) treated the GCs of the Galaxy as a tracer of its chemical evolution and concluded that the Galaxy was formed by an inhomogeneous collapse.

Kinman (1959) demonstrated that the GCS of our Galaxy as a whole has a small net rotation and Frenk \& White (1980) suggested that the difference in rotational velocity between the F-type (metal-poor) and G-type (metal-rich) GC groups is only marginally significant in our Galaxy.
Huchra et al. (1982) also discovered a rotating system of M31 GCs.
Zinn (1993) pointed out that inner "old halo" clusters have rapid rotation, small velocity dispersion, and flattened orbits and suggested that these properties are evidence of the dissipational collapse formation of our Galaxy.

Minniti (1996) studied the kinematics of GCs and stars in the Galactic bulge and concluded that the metal-rich GCs and stars in the Galactic bulge have similar kinematic properties to the inner "old halo" population (cf. Zinn 1993) and that the Galactic bulge was formed by dissipational collapse.

Recently, Chiba \& Beers (2000) discussed a new scenario for the formation of our Galaxy from studying the proper motions of metal-poor halo stars in the solar neighborhood obtained by HIPPARCOS.

In studying the global dynamic evolution of galaxies, it is advantageous to use the kinematic properties of the GCSs of those galaxies, because most GCs should have been formed before, during, or immediately after the collapse of their host galaxies (Fall \& Rees 1985). Those GCs, therefore, can give us information on the dynamic evolution that occurred during those epochs.

We discuss the dynamic formation scenario of disk galaxies using the kinematic property of the GCS, especially the relation between metallicity and the ratio of the rotational velocity to the velocity dispersion.

We construct simple models to simulate the evolution of the chemo-dynamic properties of the GCSs of disk galaxies in Section 2 and compare the observed properties of the GCSs of our Galaxy and M31 with those derived from these simple models in Section 3. In Section 4, we discuss the importance of observing GCSs in S0 and Sa type galaxies. Finally, in Section 5, we try to recompile the GCS data of M81 and propose that the chemo-dynamic evolution of the M81 system may differ from those of the Galaxy and M31. We then conclude that we must investigate the detailed kinematics of the M81 GCS, because the quality and quantity of the data on M81 GCs is poor.

\section{Simple Models for the Dynamic Evolution of GCS}

\subsection{Model Construction}

In this section, we investigate the time variation of the rotational velocity of the collapsing protogalactic gas of the disk galaxy from which GCs were formed, and search for a relation between the metallicity $Z$ and the normalized specific rotation \vsig\ of GCS, where $<v_{\phi}^{obs}(Z)>$ and $\sigma_{v}(Z)$ are the mean rotational velocity and the velocity dispersion of the GCS, respectively.

In order to make the dynamic models of disk galaxy formation treatable, we assume that the protogalaxies underwent homologous dynamic contraction along a family of self-consistent density-potential pair solutions of the Poisson equation by conserving its angular momentum.

The axisymmetric models adopted in this paper are those derived by Miyamoto \& Nagai (1975) and their density $\rho$ and the gravitational potential $\Psi$ are given by

\begin{equation}
\rho(r,z;a(t),b(t)) = \frac{b^2M_{gal}}{4\pi}
\frac{ar^2 +(a+3\sqrt{z^2 +b^2})(a+\sqrt{z^2 + b^2})^2}
{\{r^2 + (a+\sqrt{z^2+b^2})^2\}^{5/3}(z^2+b^2)^{3/2}}\;\mbox{,}
\end{equation}
\begin{equation}
\Psi(r,z;a(t),b(t)) = -\frac{GM_{gal}}{\sqrt{r^2 +(a + \sqrt{z^2 + b^2})^2}}\;\mbox{,}
\end{equation}

where $M_{gal}$ is the total mass of the galactic disk and the parameters $a(t)$ and $b(t)$ are the time-dependent scale lengths along the r and z directions, respectively.
Solutions (1) and (2) for the density-potential pair satisfy the Poisson equation given by

\begin{equation}
\nabla ^2\Psi = 4\pi G\rho \;\;\mbox{.}
\end{equation}

For simplicity, we assume that $b/a = 0.2$ at all times, which is acceptable for typical disk galaxies.
The contraction profile of the protogalaxies is then expressed by the variation of the scale length $a(t)$ as a function of time $t$.

Furthermore, we assume that the total angular momentum $J_{\phi}$ of the protogalaxies is conserved during the collapse phase and its direction is perpendicular to the galactic disk.
The expression for the total angular momentum is given by

\begin{equation}
J_{\phi} = \int_{0}^{R}2\pi v_{\phi}(r,t)\Sigma (r,t)r^2dr \;\;\mbox{,}
\end{equation}

where $v_{\phi}(r,t)$ and $\Sigma (r,t)$ are the rotation velocity parallel to the galactic disk and the projected surface density defined by

\begin{equation}
\Sigma (r,t) = \int_{-H}^{H}\rho(r,z,t)dz \;\;\mbox{,}
\end{equation}

where $\rho (r,z,t)$ is the volume density of the galaxy. For practical calculation purposes, we adopt the radius $R = 4a$ and the half thickness $H = 4b$. It has been shown that $\sim 10^{-6}\rho _{c}$ at these radius and half thickness defined by $R$ and $H$, where the $\rho _{c}$ is the central mass density of a disk galaxy.

Since a flat rotation curve is a typical property of disk galaxies, we assume that

\begin{equation}
<v_{\phi}(r,t)> = v_{flat}(t) \;\;\mbox{.}
\end{equation}

This requires that when the density profile varies with the time-dependent scale radius $a(t)$, the angular momentum distribution is adjusted to make up the flat $v_{\phi}$, whereas the total angular momentum of the galaxy is conserved.

The flat rotation velocity is then given by

\begin{equation}
v_{flat}(t) = \frac{J_{\phi}}{I_{R}(t)}\;\mbox{,}
\end{equation}

where the moment of inertia $I_{R}(t)$ is defined by

\begin{equation}
I_R(t) = \int_{0}^{R}2\pi \Sigma (r,t)r^2dr\;\mbox{.}
\end{equation}

For simplification, in this paper, we assume that the spatial distribution, angular momentum, total energy, and flat property of the rotation curves of these GCs have been frozen and conserved since the time of GCS formation. Furthermore, we assume that once GCs form in the collapsing protogalactic gas, they are disconnected from the collapsing gas system.
If the kinematical energy of GCS were not conserved, the $ <v_{rot}^{obs}(Z)>$ should be larger than that of conserved case for metal-poor clusters.
However the $\sigma_{v} $ also should be larger, so the effect of the dissipation for \vsig \ can be negligible. 
 
Under these assumptions, the mean rotational velocity $<v_{\phi}^{t_1} ({\mbox{\boldmath $r$}},Z)>$ of GCs observed at present($t=t_1$) would be equivalent to the mean rotational velocity $v_{flat}(t)$ during the formation of those GCs. The age-metallicity relation that translates $t$ to $Z$ is discussed in Section 2.3.

Other velocity components $v_r^{t_1}({\mbox{\boldmath $r$}},Z)$ and
$v_z^{t_1}({\mbox{\boldmath $r$}},Z)$ depend on $<v_{\phi}^{t_1} ({\mbox{\boldmath $r$}},Z)>$ and the present gravitational potential $\Psi(r,z;a(t_1),b(t_1))$.
If the disk galaxies are well virialized, those velocity components are given by

\begin{equation}
v_r^{{t_1} ^2}({\mbox{\boldmath $r$}},Z) = v_z^{{t_1} ^2}({\mbox{\boldmath $r$}},Z) = \Psi _{1} - \frac{J_{\phi}^2}{2I_R^2(t)} \;\;\mbox{.}
\end{equation}

We then derive the formulation for the observational, normalized specific rotation \vsig\ of an disk galaxy from those three velocity components and the inclination $i$ of the disk projected onto the celestial sphere.
By contrast, the observed \vsig\ of our Galaxy is derived by assuming that the observer is at the solar radius (see Appendix).

The parameters adopted in this paper are:
the scale length $a_1$ of a typical disk galaxy at present is $a_1 \sim 6.25 $kpc, hence $R_1 = 25 $kpc (e.g. Helmi \& White 1999, Patsis \& Grosb\o l 1996), the initial($t = t_0$) disk radius of a galaxy is $R_0 = 3R_1$ (e.g., Rees \& Ostriker 1977), and the present flat rotation velocity is $v_{rot}(t_1) \sim 250 km\; s^{-1}$ (e.g., Rubin \& Ford 1970, Burton \& Gordon 1978).

\subsection{Collapse Profile of Scale Radius $a$}

It is not straightforward to solve the time variation of the scale radius a(t) of the gas self-consistently.
Instead, we adopt three models for describing the collapse profiles of protogalaxies for consideration.

(1) Exponential collapse: In this model, we assume that the scale size of the protogalaxy decreases exponentially with time, and that $a(t)$ can be written in the form

\begin{equation}
a(t) = a_0 \exp\left(-\frac{t}{t_{col}}\right)\;\mbox{,}
\end{equation}

where $t_{col}$ is the timescale of collapse and $a_0$ is the initial size of the protogalaxy.
Of the three models, this represents a moderate collapse.

(2) Delayed collapse: In this model, the size of the protogalaxy decreases gradually with time initially and then, after some epoch, it decreases more rapidly with time.
For our Galaxy, based on Zinn(1993), Chiba \& Beers (2000), this scenario assumes that the halo was built up by numerous fragments falling into the Galaxy (e.g., Searle \& Zinn 1978) and that at some time the halo collapsed rapidly (e.g., ELS).
We assume that the slow collapse process was dominant until $t_{trans}$ and that the rapid collapse process became dominant after $t_{trans}$, namely

\begin{equation}
\begin{array}{ccll}
a(t) & = & a_0 \exp\left(-\frac{t}{t_{col,1}}\right) & (t < t_{trans})\\

a(t) & = & a_0 \exp\left(-\frac{t_{trans}}{t_{col,1}} + \frac{t_{trans}}{t_{col,2}}\right) \exp\left(-\frac{t}{t_{col,2}}\right) & (t > t_{trans})\;\mbox{,}
\end{array}
\end{equation}

where $t_{col,1}$ and $t_{col,2}$ are the long and short time scales (i.e., $t_{col,1} >  t_{trans} > t_{col,2}$), characterizing the two-stage collapse taking place before and after the transition epoch $t_{trans}$.

(3) Early collapse: In this model, the protogalaxy experiences a rapid initial collapse, followed by a slow collapse.
The equation describing this collapse profile is the same as for the delayed collapse model; however, the time scale of the first collapse, $t_{col,1}$, is smaller than that of the second collapse, $t_{col,2}$.
Although this model is not supported by theoretical or observational evidence, we examine this model as an extreme case, in contrast to the delayed collapse model.

The collapse profiles describing the time variation of $a(t)$ for the protogalaxy in these three collapse models are shown in Figure 1.

\subsection{Age Metallicity Relation}

This subsection discusses the temporal evolution of metallicity, that is, the so-called age metallicity relation (AMR) of disk galaxies under the assumption that the metallicity of a GC traces the chemical evolution of the collapsing protogalaxy.
The age of a galaxy is defined as the elapsed time from the start of the initial collapse of the galaxy.
The initial mass function and rate of star formation are probably dependent on the density and pressure of the collapsing gas.
Therefore, the AMR might well be modified according to the phase of star formation.
For simplification, however, we assume that the AMR is independent of the dynamic effect.

We adopt a simple closed-box model for the chemical evolution of disk galaxies. This simple closed-box model is based on Hartwick's (1976) concept, introduced as a simple model for the chemical evolution of the Galactic halo.
We assume that the GCs were formed from the collapsing gas and that the GCs conserve their initial chemical composition.
Under this assumption, the chemical abundance in GCs should reflect the chemical abundance in the collapsing halo gas during the formation of those GCs.

This simple closed-box model adopts the Schmidt Law $\psi _{sf}(t) \propto f_g^n(t)$ for the star formation rate, where $f_g(t)$ is the gas fraction of the galaxy.

The AMR of a disk galaxy is obtained by coupling the basic equations for chemical evolution (Tinsley 1980) with the simple closed-box model.
It is given by

\begin{equation}
\log (Z/Z_{\odot}) = \log (y/Z_{\odot}) + \log (t/t_{sf}) \;\;\;(n=1)\;\mbox{,}
\end{equation}
or
\begin{equation}
\log (Z/Z_{\odot}) = \log (y/Z_{\odot}) + \log \{\log(1 + t/t_{sf})\}\;\;\;(n=2)\;\mbox{,}
\end{equation}

where $Z_{\odot}$ is the metallicity of the solar neighborhood, $t$ is the age of the galaxy since the onset of the initial collapse, $t_{sf}$ is the time scale of the active star formation period, and $y$ is the yield of newly created elements. These equations are quoted from Arimoto, Yoshii \& Takahara (1992).
For our Galaxy, the observation implies $y \sim 0.1Z_{\odot}$ in the halo.
In this work, we assume that nearby disk galaxies are the same age as the Galaxy, so they have the same star formation rate as the Galaxy.

\subsection{Implication of Simple Model}

Figure 1 shows the collapse profiles of the scale radius $a(t)$ for the three models studied in this paper. Figure 2 shows the evolution of the normalized specific rotation \vsig\ of the disk galaxy as a function of metallicity $\log Z/Z_{\odot}$ for each of the three collapse models for $n = 1$.
The exponential and delayed collapse models both show rather an abrupt increase in \vsig\ at $\log Z/Z_{\odot} \sim -1$ for $t_{col}/t_{sf} = t_{trans}/t_{sf} = 1$ and at $\log Z/Z_{\odot} \sim -0.0$ for $t_{col}/t_{sf} = t_{trans}/t_{sf} = 10$. The epoch of this abrupt increase in \vsig\ depends directly on the value of $t_{sf}$.
For the early collapse model, \vsig\ increases gradually in contrast to the exponential and delayed collapse models.
Thus, we can discriminate between the early collapse model and the other two models in terms of the increase in \vsig.

Figure 3 shows a relation similar to that of Figure 2, but for $n = 2$.
Again, we see an abrupt increase in \vsig\ at some $\log Z/Z_{\odot}$ for the exponential and delayed collapse models.
In these two models, the abrupt increase in \vsig\ is a universal property in contrast to the early collapse model.
Therefore, we can easily see whether the early collapse model adequately describes the dynamic collapse history of disk galaxies, by examining the observed \vsig\ of the GCSs in disk galaxies.

From Figures 2 and 3, we conclude that when $n$ and $t_{col}/t_{sf}$ or $t_{trans}/t_{sf}$ are fixed in a model, the profiles of the increase in \vsig\ are very similar. In summary,

(1) The evolution of the normalized specific rotation as a function of the metallicity of GCS, described by the relation \vsig\, can be used as a diagnostic tool to identify the collapse history of the halos of disk galaxies.

(2) When the Schmidt Law Index $n$ is fixed, the $Z$-dependence of the ratio of \vsig\ is sensitive to the ratio $t_{col}/t_{sf}$, but is insensitive to the individual value of $t_{col}$ or $t_{sf}$ for a common ratio of $t_{col}/t_{sf}$.

\section{The Observational Relation between $<v_{rot}^{obs}>/\sigma_{v}$ \ and [Fe/H]}

\subsection{Our Galaxy}

We now compare the observed kinematic properties of GCSs with those predicted from the simple models discussed in Section 2.
The data of the GCS of our Galaxy, which are used to derive the \vsigmr\ presented in this paper, are based on the most recent catalogue of Harris (1996).
In this catalogue, only the heliocentric velocity $v_h$ is available. First, we correct $v_h$ to $v_{rad} = v_h + v_{\odot}\sin l \cos b$, where $v_{\odot}$ is the velocity of the LSR about the galactic center, and $l$ and $b$ are Galactic longitude and latitude, respectively.
Therefore, we regard the mean radial velocity as the rotational velocity. In order to derive it, we divide the Galactic disk into two regions: one includes longitudes with $0^{\circ} < l < 180^{\circ}$ and the other includes longitudes with $180^{\circ} < l < 360^{\circ}$.
Then, we calculate the mean of radial velocity for each of the regions, $<v_{rad}^{+}>$ and $<v_{rad}^{-}>$ and regard $<v_{rad}^{+}>-<v_{rad}^{-}>$ as $<v_{rot}^{obs}>$.
We use [Fe/H] as an indicator of the metallicity $Z$.

We adopt the "{\it{running mean}}" method to evaluate \vsig\ from a limited number of available samples, and hence are interested only in global properties of the derived \vsig\ .

The general properties of the derived \vsigmr \ of the GCS of our Galaxy (Figure 4) are as follows:

(1) The GCs having ${\mbox{[Fe/H]}} < -1.0$ show only a modest net rotation $<v_{rot}>/\sigma \sim 0.5$.

(2) The rotational velocity of GCs increases rather abruptly from ${\mbox{[Fe/H]}} \sim -1.0$ to ${\mbox{[Fe/H]}} \sim -0.7$.

In our Galaxy, the metallicity distribution of the GCS shows the presence of two distinct components (Zinn 1985). This bimodality suggests that two different formation processes were involved.
It appears that one of the populations was mixed with the other at [Fe/H]$\geq -1.0$.
Therefore, the assumption that the GCS was formed from the collapsing proto-Galaxy might not hold in that range of metallicity.

This paper examines the chemical evolution of GCs that formed during the halo collapse phase with various collapse profiles, but it does not consider the deviation of the age-metallicity relation of the disk population of GCs from that of the halo population. Our simple model reproduces the general increasing trend of \vsig\ with increasing [Fe/H], but the low values of \vsig\ were reproduced at $-1.8 < $[Fe/H]$ < -1.0$.
This might be due to a major merging episode that took place during this epoch, since major merger events can increase $\sigma_{v}$ temporally and decrease net $<v_{rot}^{obs}(Z)>$ to lower \vsig\ .

\subsection{M31}

M31, the nearest spiral galaxy, has a similar mass ($1.8 \times 10^{11}M_{\odot}$; Rubin \& Frod 1970) and morphological type (Sb; de Vaucouleurs et al. 1991) to our Galaxy, and is a natural test bed to apply our scheme for studying chemo-dynamic evolution using its GCS.

We recompiled the most recent GCS catalogue of M31 (Barmby et al. 2000) to derive \vsigmr .
We regard the mean radial velocity as the rotational velocity. In order to derive it, we divide the M31 disk into two regions: one has a direction along with major-axis of M31 disk from center of M31 and the other has the other direction.
Then, we calculate the mean of radial velocity for each of the regions, $<v_{rad}^{+}>$ and $<v_{rad}^{-}>$ and regard $<v_{rad}^{+}>-<v_{rad}^{-}>$ as $<v_{rot}^{obs}>$.

By adopting the inclination of M31, $77^{\circ}$ (e.g., Rubin \& Ford 1970), we derived Figure 5 for M31, in the same manner as Figure 4.
The overall property of \vsigmr\ for M31 is similar to that for our Galaxy.
By comparing the observed \vsigmr s of the GCSs of our Galaxy and M31 with the results of the simple model discussed in Section 2, we arrived at the following two conclusions:

(1) The contraction of the protogalaxy of M31 followed a collapse profile similar to all collapse models.

(2) The ratio of the collapse time scale to the star formation time scale should have been $t_{col}/t_{sf} \sim 3-11$ for the Galaxy, while it should be $t_{col}/t_{sf} \geq 7$ for M31, when the power index of the Schmidt Law for star formation is $n=1$.  For $n=2$, it should be $t_{col}/t_{sf} \geq 600$ for the Galaxy and $t_{col}/t_{sf} \geq 1.0\times10^4$ for M31. However, we notice that $t_{col}/t_{sf} \geq 100$ may be unusual value for that ratio. 

The range of $t_{col}/t_{sf}$ is derived considering uncertainties of our model and the observed \vsigmr , i.e. if a model line with a $t_{col}/t_{sf}$ is covered error bars of metal-rich GCs' data points, this $t_{col}/t_{sf}$ is accepted.
We regard the range of $-1.0 < $[Fe/H]$ < -0.3$ and $-0.6 < $[Fe/H] as metal-rich GCs' data points for our Galaxy and M31, respectively. 
For our Galaxy, we leave the point of [Fe/H]$= -0.2$ out of consideration, because it should be unexpected point for our model.  

\section{Effects of Late Mergers on \vsigmr}

Recently, the hierarchical clustering model has been discussed as a standard galactic formation model (e.g., Katz \& Gunn 1991).
Merger events probably played a considerable role in triggering GCs formation (Ashman \& Zepf 1992).
There is evidence that some elliptical galaxies have undergone mergers (e.g., Zepf \& Ashman 1993).
Zepf \& Ashman (1993) predicted that the GCSs of some elliptical galaxies have two peaks in their metallicity distribution.
Our simple model assumes that galaxies were formed through the monolithic collapse of primordial gas.
Our dynamical model is compatible with a hierarchical clustering model, if the clustering gas clumps are numerous and have small scale.
However, if the clustering of gas clumps involves a major merger event, our model may not adequately describe the dynamic and chemical evolution of such a system.

By contrast, for those dissipationless merger events that did not trigger the formation of GCs, only the kinematics of GCs were affected. In fact, Kissler-Patig \& Gebhardt (1998) found evidence for dissipationless merger events in the globular clusters kinematics of M87. Some S0 and Sa galaxies might also have undergone dissipationless merger events (e.g., Jore, Broeils \& Haynes 1996, Mihos, Walker \& Hernquist 1995 ).
The occurrence of similar dissipationless merger events is conceivable in disk galaxies.
If dissipationless merger events played a significant role in disk galaxies, \vsigmr\ cannot be predicted using the simple models introduced in Section 2.
In other words, if the formation of the GCS is affected by dissipationless merging events, we will find that its kinematical properties are more complicated and will not show the rather simple relations observed in our Galaxy and M31 (cf. Figures 4 and 5).

We need to investigate the relation between the kinematics and metallicity for the GCSs of early-type disk galaxies to see the extent of the role of dissipationless merging events in the formation of GCSs.

\section{A Preliminary Assessment of the \vsigmr\ for M81}

Elaborate observational studies of the GCSs of disk galaxies have been carried out during the last decade.
After the Galaxy and M31, the GCS of M81 is one of the most extensively observed (Georgiev et al. 1992, 1993, Perelmuter \& Racine 1995, Perelmuter, Brodie \& Huchra 1995).
Perelmuter, Brodie \& Huchra (1995) published kinematical data of the GCs of M81.
Even so, at present only 25 M81 GCs are available for the study of their chemo-dynamical properties.
Figure 6 shows the \vsigmr\ for M81, indicating a increasing
\vsig\ at all [Fe/H] similar to our Galactic and M31 GCS. M81 likely underwent a significant interaction with M82 and NGC 3077, the massive member galaxies of the M81 group.
The evidence for such an interaction is the neutral hydrogen stream between those galaxies that has been confirmed by 21cm HI emission studies (van der Hulst 1979, Yun et al. 1993). Therefore, the kinematics of the M81 GCS might have been modified by tidal effects more conspicuously than in the Galaxy or M31.

At present, our study is limited by the paucity of available data on the GCS of M81.
Since the expected number of GCs in M81 is more than 150, we are planning an extensive study of the properties of the GCS of M81 to constrain the scenario of M81 formation.

\section{Conclusion}

We made a series of chemo-dynamic model calculations to predict the relation for GCSs under the assumption that GCSs are formed before or during the collapse of protogalactic gas clouds.

Three collapse profiles, e.g., exponential collapse, delayed collapse, and early collapse, are assumed in our simple models.
The power index $n$ of the Schmidt Law for the star formation process, the dynamic collapse time scale $t_{col}$, and the star formation time scale $t_{sf}$ are the free parameters chosen in this study.

We analyzed the \vsigmr s of our Galaxy (Figure 4) and M31 (Figure 5).
The observed relations for these two disk galaxies of similar mass and morphological type are similar.
They exhibit a rather sudden increase in \vsig\ at [Fe/H]$\sim -1.0$ for the Galaxy and at [Fe/H]$\sim -0.6$ for M31.
This is consistent with chemo-dynamic model predictions of the formation of the GCSs of these galaxies. However we have to notice that it is difficult to choose the parameter set ($t_{col,1}$,$t_{col,2}$) for fitting the model with data points for early collapse model.
It also suggests that the GCSs of these galaxies were formed in a dynamic collapse, where the collapse time scale is as large as, or larger than, the star formation time scale, if the star formation process is proportional to the protogalactic gas density, namely $n=1$.
The collapse time scale should be at least an order of magnitude larger than the star formation time scale, if the star formation process took place more efficiently in an enhanced density region, as is the case for $n=2$.

We noticed a difference in the [Fe/H] value at which the abrupt increase of \vsig\ takes place between the Galaxy and M31.
By comparing these observed curves with the results of simple models, we derived ratios of the collapse time scale to the star formation time scale for the Galaxy and M31 of $t_{col}/t_{sf} \sim 3-11$ and $t_{col}/t_{sf} \geq 7$, respectively, when the power index of the Schmidt Law for star formation is $n=1$.
For $n=2$, those ratios are $t_{col}/t_{sf} \geq 100$. This might be a evidence that the $n=2$ model is ruled out, however we have to construct more realistic chemo-dynamical evolution model for discussing whether it is ruled out or not.   
These differences in the [Fe/H] value do not reflect a significant differences in the $t_{col}/t_{sf}$. 
We suggest that the difference of the \vsigmr \ between galaxies should be a probe for studying the differences of formation scenarios of those galaxies, although we cannot restrict the formation scenario of a galaxy from it.

For the GCSs of both our Galaxy and M31, we see that the value of \vsig\ is $\sim 0$ at $-1.5<$[Fe/H]$<-0.8$, which is not consistent with the models.
Since we assumed that the protogalactic gas was monolithic in our models, \vsig\ may have been affected by merger events during the epoch that correspond to that metal range.

GCs are useful probes for investigating the chemo-dynamic properties of host galaxies.
New observations of the GCSs of nearby disk galaxies with 8-10m class telescopes will provide fruitful suggestions on the history of galaxy formation.

\appendix
\section{Scheme for deriving the rotational velocity $<v_{rot}^{obs}(Z)>$ and Velocity dispersion $\sigma _v(Z) $}

We describe the scheme we used to derive the average rotational velocity $<v_{rot }^{obs}(Z)>$ and
the velocity
dispersion $\sigma _v (Z)$ from the velocity components ($v_r^{t_1}
({\mbox{\boldmath $r$}},Z)$, $v_{\phi}^{t_1} ({\mbox{\boldmath
$r$}},Z)$, $v_z^{t_1} ({\mbox{\boldmath $r$}},Z)$)(cf. Figure 7) observed 
line-of-sight velocity $v^{obs}_{l.o.s}({\mbox{\boldmath $r$}},Z)$ of GCs.

 The observed line-of-sight velocity $v^{obs}_{l.o.s}({\mbox{\boldmath $r$}},Z)$ of GCs is the summation of the projected components of the actual 3D velocity components ($v_r^{t_1}
({\mbox{\boldmath $r$}},Z)$, $v_{\phi}^{t_1} ({\mbox{\boldmath
$r$}},Z)$, $v_z^{t_1} ({\mbox{\boldmath $r$}},Z)$)(cf. Figure 7 for a disk galaxy and Figure 8 for our Galaxy) and as is expressed by

\begin{equation}
v^{obs}_{l.o.s}({\mbox{\boldmath $r$}},Z) = v_{r}^{t_1}({\mbox{\boldmath $r$}},Z)\sin\theta \cos i  + v_{\phi }^{t_1}({\mbox{\boldmath $r$}},Z)\cos \theta \cos i + v_{z}^{t_1}({\mbox{\boldmath $r$}},Z)\sin i \;\mbox{,}
\end{equation}

for a disk galaxy, and

\begin{equation}
v^{obs}_{l.o.s}({\mbox{\boldmath $r$}},Z) = v_{r}^{t_1}({\mbox{\boldmath $r$}},Z)\sqrt{1-\frac{R_{\odot }^2}{r^2}\sin ^2 l}\frac{r^{\prime }}{\sqrt{r^{\prime 2} + z^2}}  + v_{\phi }^{t_1}({\mbox{\boldmath $r$}},Z)
\frac{R_{\odot }}{r}\sin l \frac{r^{\prime }}{\sqrt{r^{\prime 2} + z^2}}
+ v_{z}^{t_1}({\mbox{\boldmath $r$}},Z)\frac{z}{\sqrt{r^{\prime 2} + z^2}}
 \;\mbox{,}
\end{equation}

for our Galaxy, where the superscript $t_1$ means that the velocity components are those observed at present($t = t_1$)(cf. section 2.1).
The average rotational velocity for GCs with metallicity $Z$ is defined as

\begin{equation}
<v_{rot}^{obs}(Z)> = \int_{{\mbox{\boldmath $r$}}}\int_{{\mbox{\boldmath $v$}}} v^{obs}_{l.o.s}({\mbox{\boldmath $r$}},Z)P(r,\theta ,z , v_{r}^{t_1}, v_{\phi }^{t_1}, v_{z}^{t_1})rdrd\theta dzdv_{r}^{t_1}dv_{\phi }^{t_1}dv_{z}^{t_1} \;\mbox{,}
\end{equation}
where $P(r,\theta ,z , v_{r}^{t_1}, v_{\phi }^{t_1}, v_{z}^{t_1})$ denotes the probability distribution function of GCs in the phase space.
We assume, for simplicity, that the probability distribution function $P(r,\theta ,z , v_{r}^{t_1}, v_{\phi }^{t_1}, v_{z}^{t_1})$ 
is spatially uniform and satisfies the symmetry requirements to suppress the net radial or vertical motion, i.e., $<v_{r}^{t_1}>=<v_{z}^{t_1}>=0$.

Then, the observed rotational velocity is given by

\begin{equation}
<v_{rot}^{obs}(Z)> = \frac{2}{\pi} <v_{\phi}^{t_1}(Z)> \cos i
\end{equation}
and the velocity dispersion is derived as

\begin{equation}
\sigma_{v}(Z) = \sqrt{<v_{rot}^{obs^{2}}(Z)> - <v_{rot}^{obs}(Z)>^2} \;\mbox{.}
\end{equation}

\figcaption[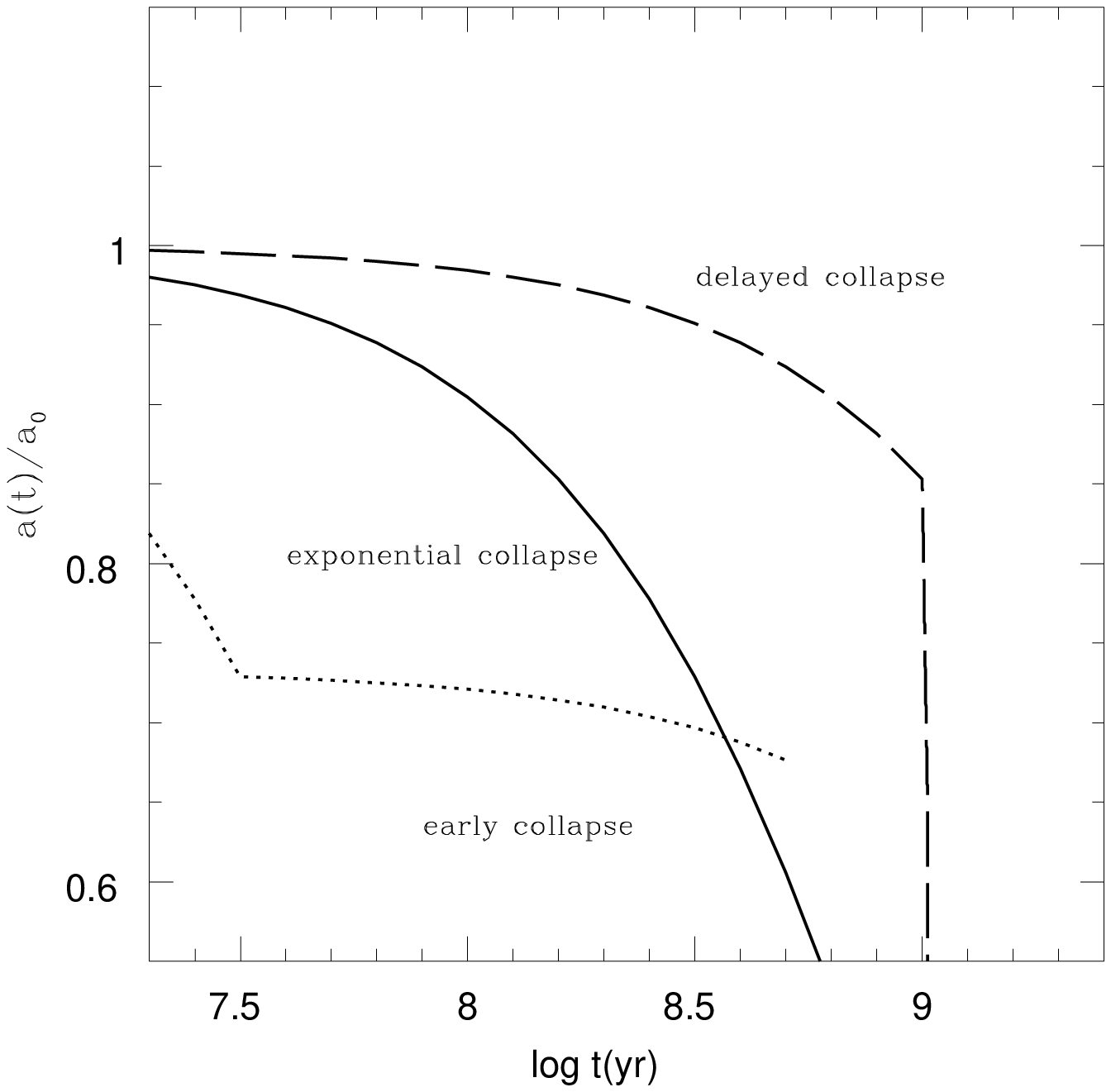]{
Collapse profiles of the scale length for three protogalaxy models.
Solid, dashed, and dotted lines show the profiles of the exponential ($t_{col} = 1 Gyr$), delayed ($t_{trans} = 1 Gyr$), and early ($t_{trans} = 0.1Gyr$) collapse models, respectively.
\label{fig1}} 
\figcaption[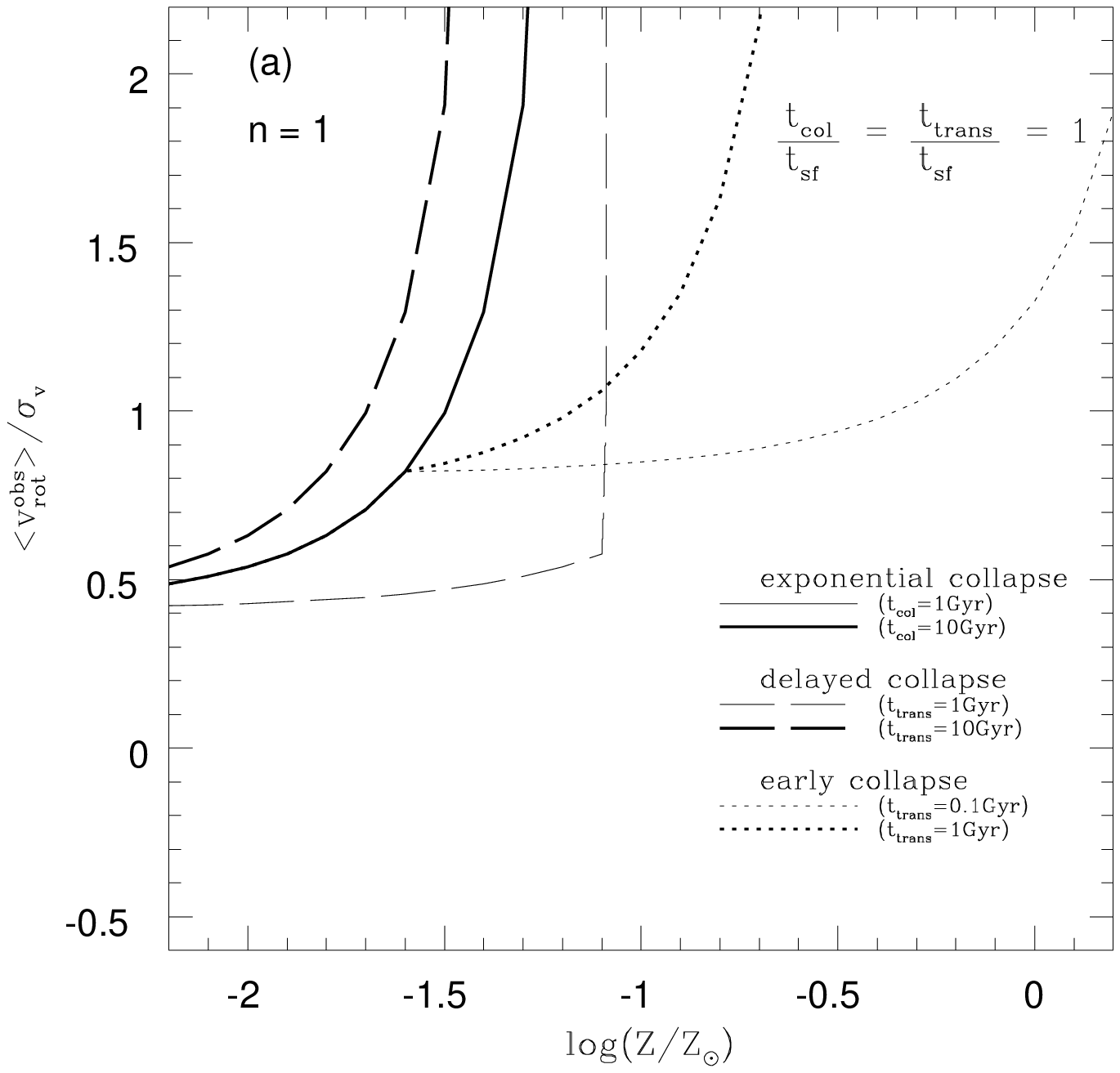, 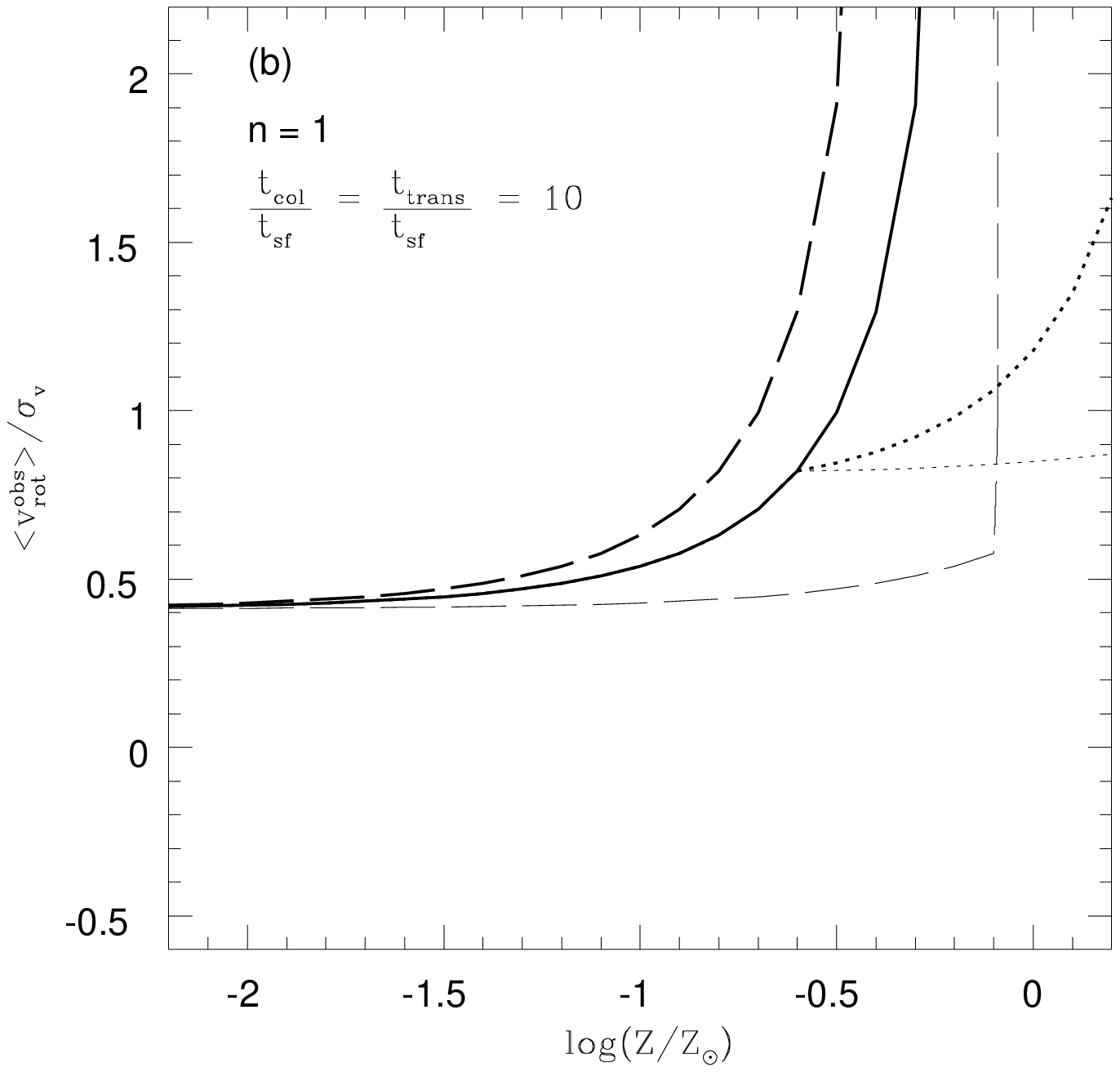]{
(a)Evolution of the normalized specific rotation \vsig\ as a function of the metallicity $log(Z/Z_{\odot}$) for models with the Schmidt Law Index $n = 1$ and $t_{col}/t_{sf} = t_{trans}/t_{sf} = 1$.
Thin and thick solid lines, which actually overlap, show the exponential collapse model for $t_{col} = 1 Gyr$ and $t_{col} = 10 Gyr$. Thin and thick dashed lines show the delayed collapse model for $t_{trans} = 1 Gyr$ and $t_{trans} = 10 Gyr$. Thin and thick dotted lines show the early collapse model for $t_{trans} = 1 Gyr$ and $t_{trans} = 10 Gyr$. The line of exponential collapse with $t_{col} = 1$Gyr overlaps the line of exponential collapse with $t_{col} = 10$Gyr.
(b)Similar diagram, but for $t_{col}/t_{sf} = t_{trans}/t_{sf} = 10$. The lines have the same meaning as in Fig. 2(a).
\label{fig2}}
\figcaption[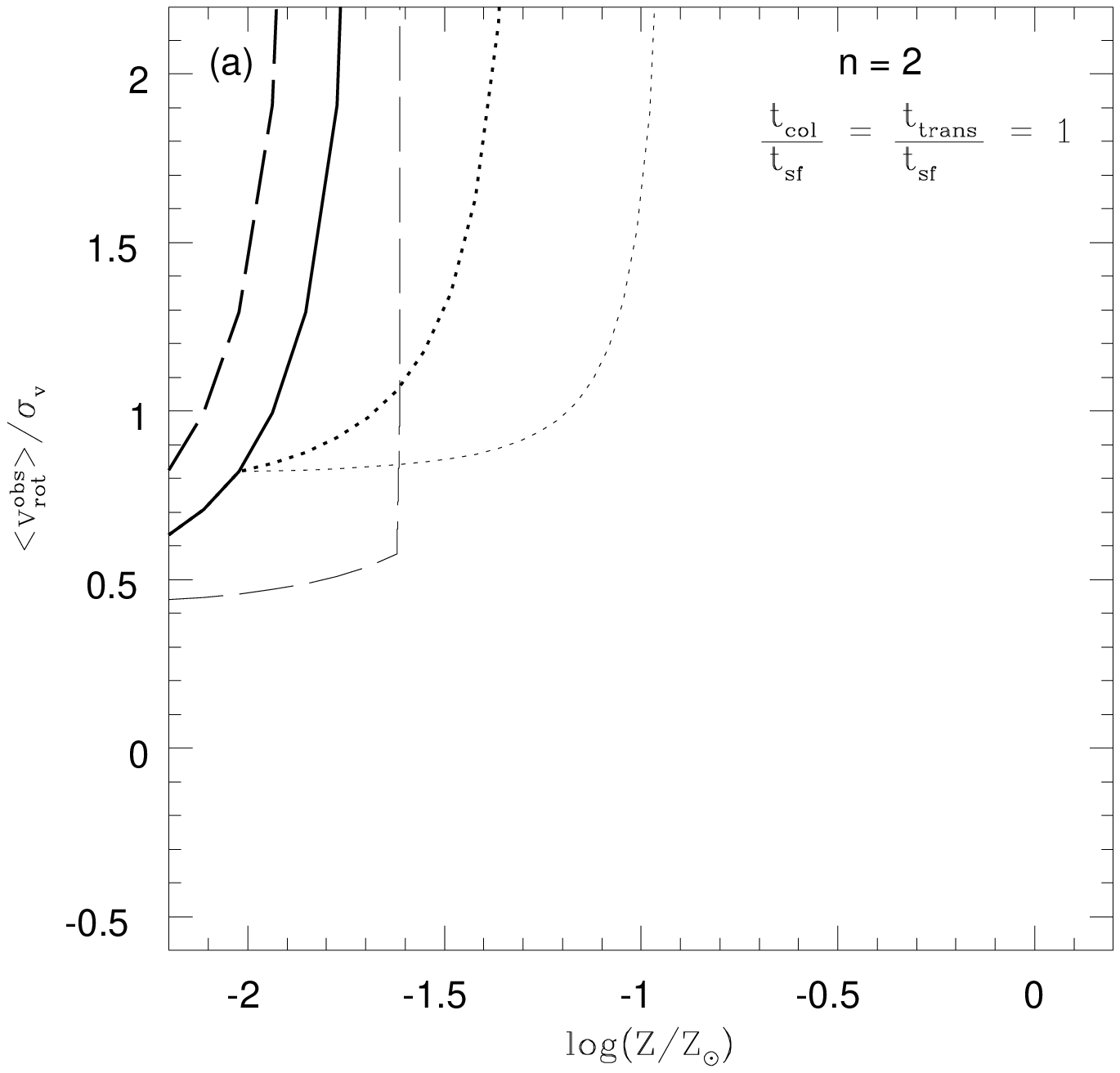, 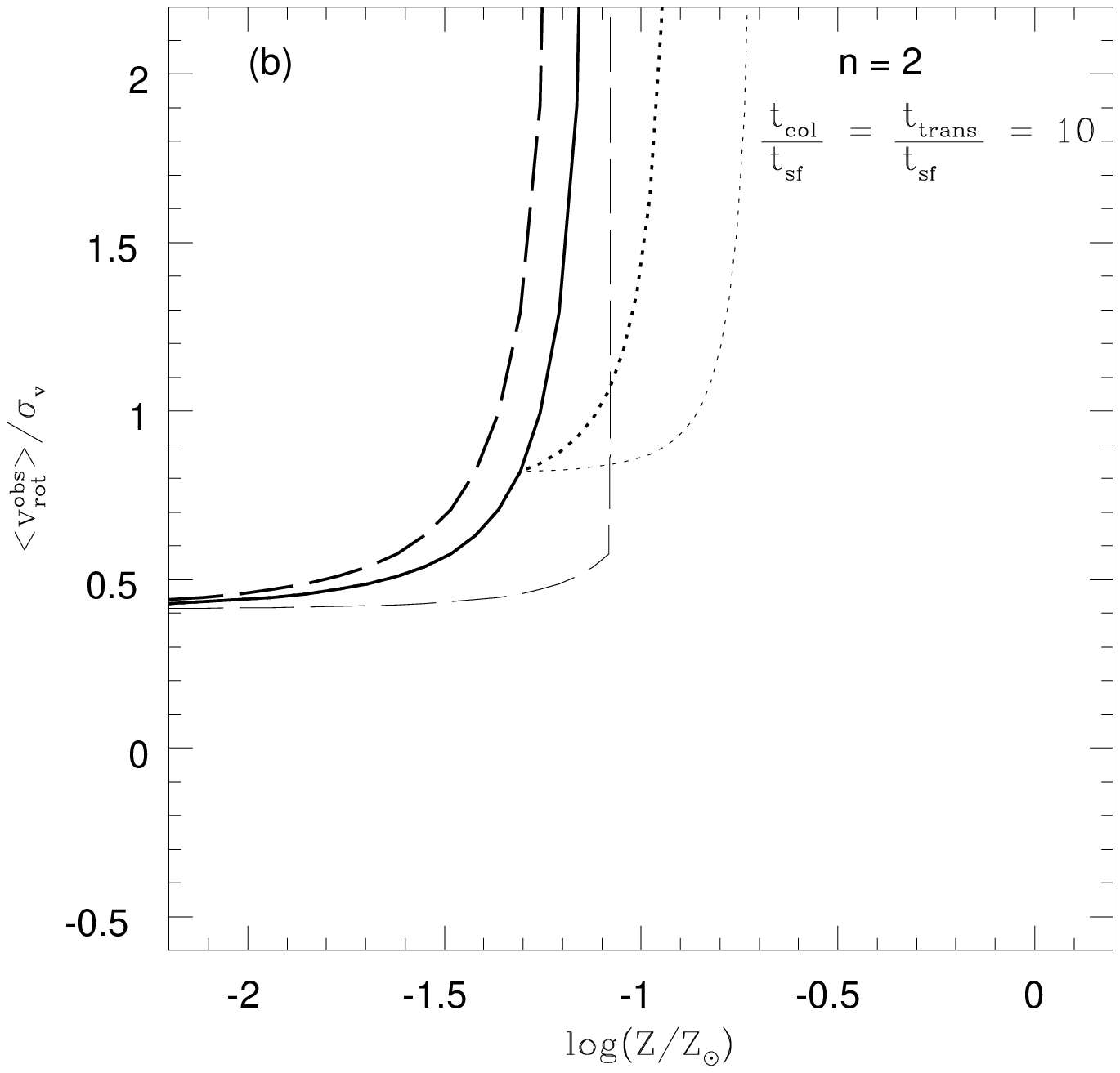]{
(a)Evolution of the normalized specific rotation \vsig\ as a function of the increasing metallicity $log(Z/Z_{\odot}$) for models with the Schmidt Law Index $n = 2$ and $t_{col}/t_{sf} = t_{trans}/t_{sf} = 1$. Other captions are the same as for Fig. 2(a). The line of exponential collapse with $t_{col} = 1$Gyr overlaps with the line of exponential collapse with $t_{col} = 10$Gyr.
(b)Similar diagram, but for $t_{col}/t_{sf} = t_{trans}/t_{sf} = 10$.
\label{fig3}} 
\figcaption[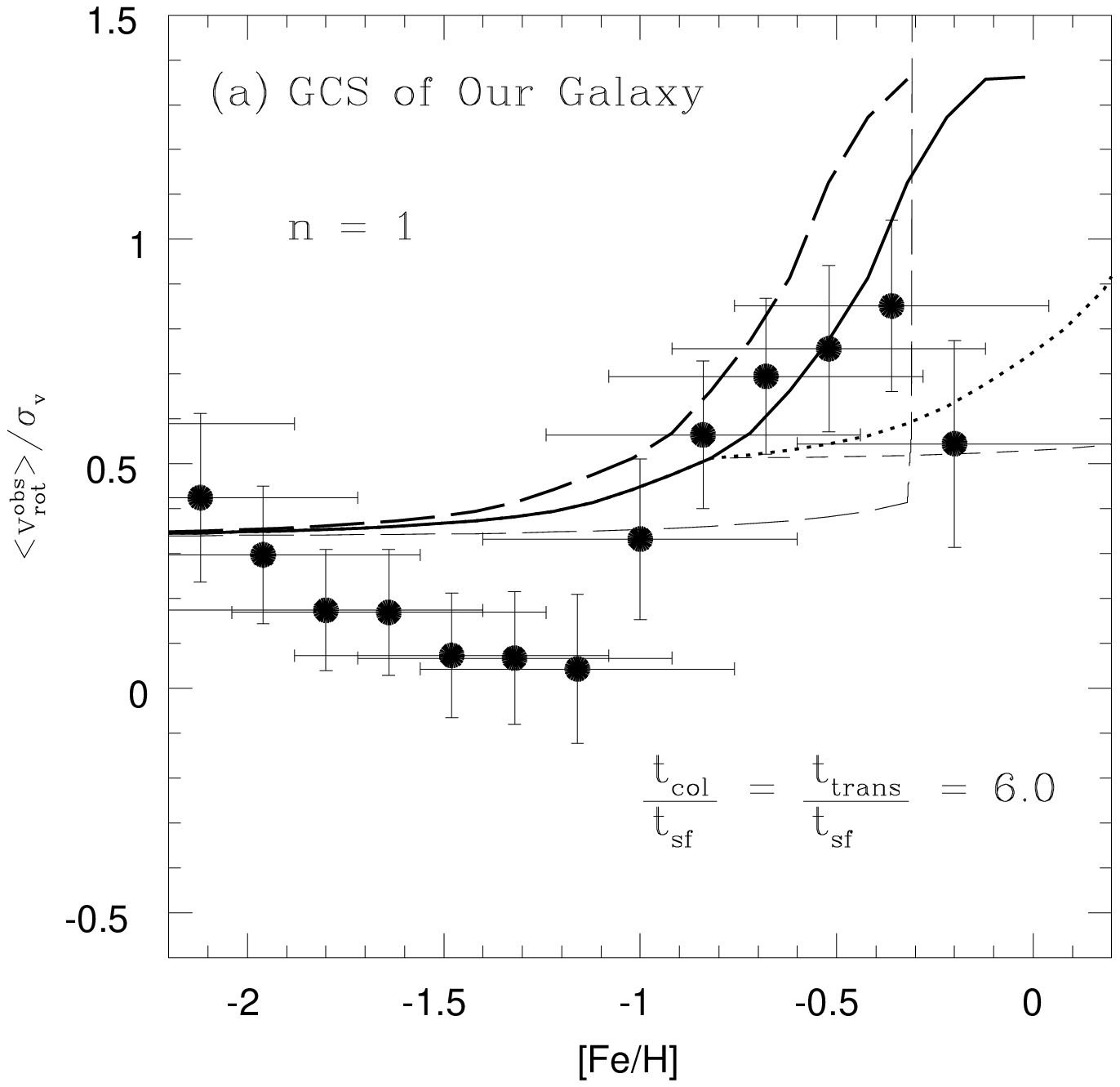, 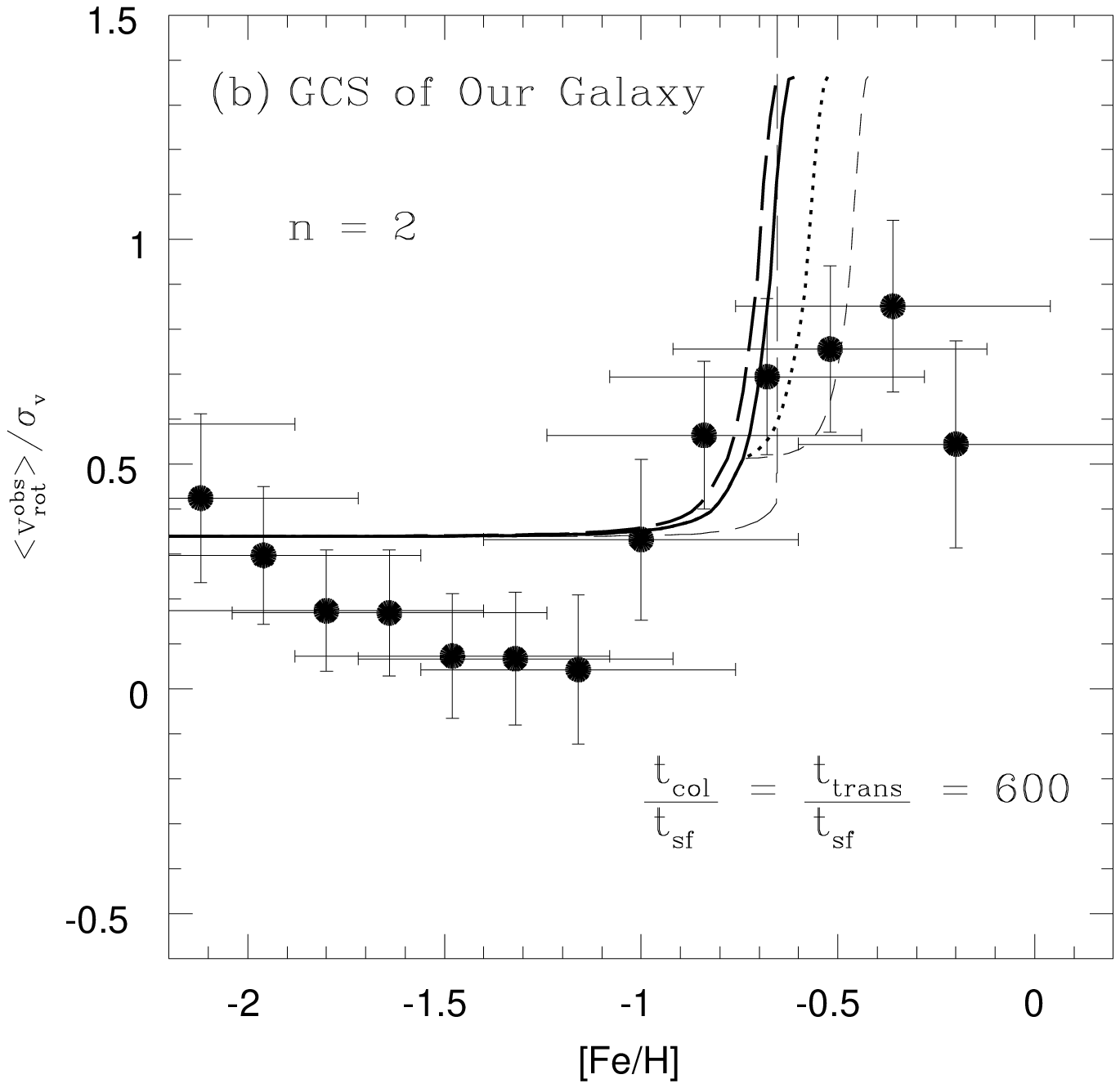]{
(a)Relation between the observed metallicity index ($[Fe/H] \equiv log(Fe/H) - log(Fe/H)_{\odot}$) and the normalized specific rotation($<v_{rot}^{obs}>/\sigma _{v}$) of the GCS of our Galaxy.
The observational data are from Harris (1996)(solid circles).
The numerical model results are as described for Fig. 2(a), but for $t_{col}/t_{sf} = t_{trans}/t_{sf} = 6.0$.
(b)Similar diagram, but for n=2 and $t_{col}/t_{sf} = t_{trans}/t_{sf} = 600$.
\label{fig4}} 
\figcaption[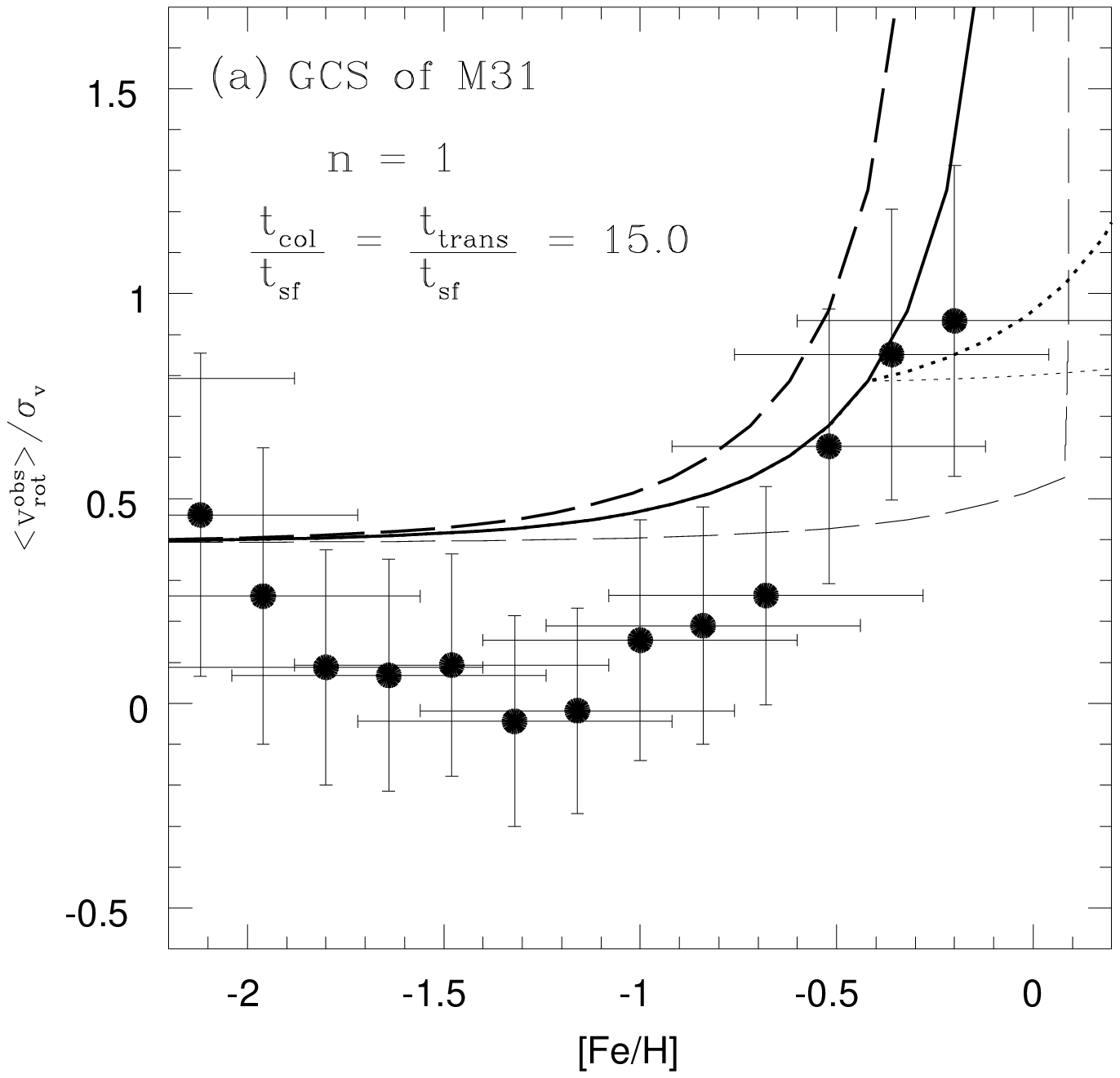, 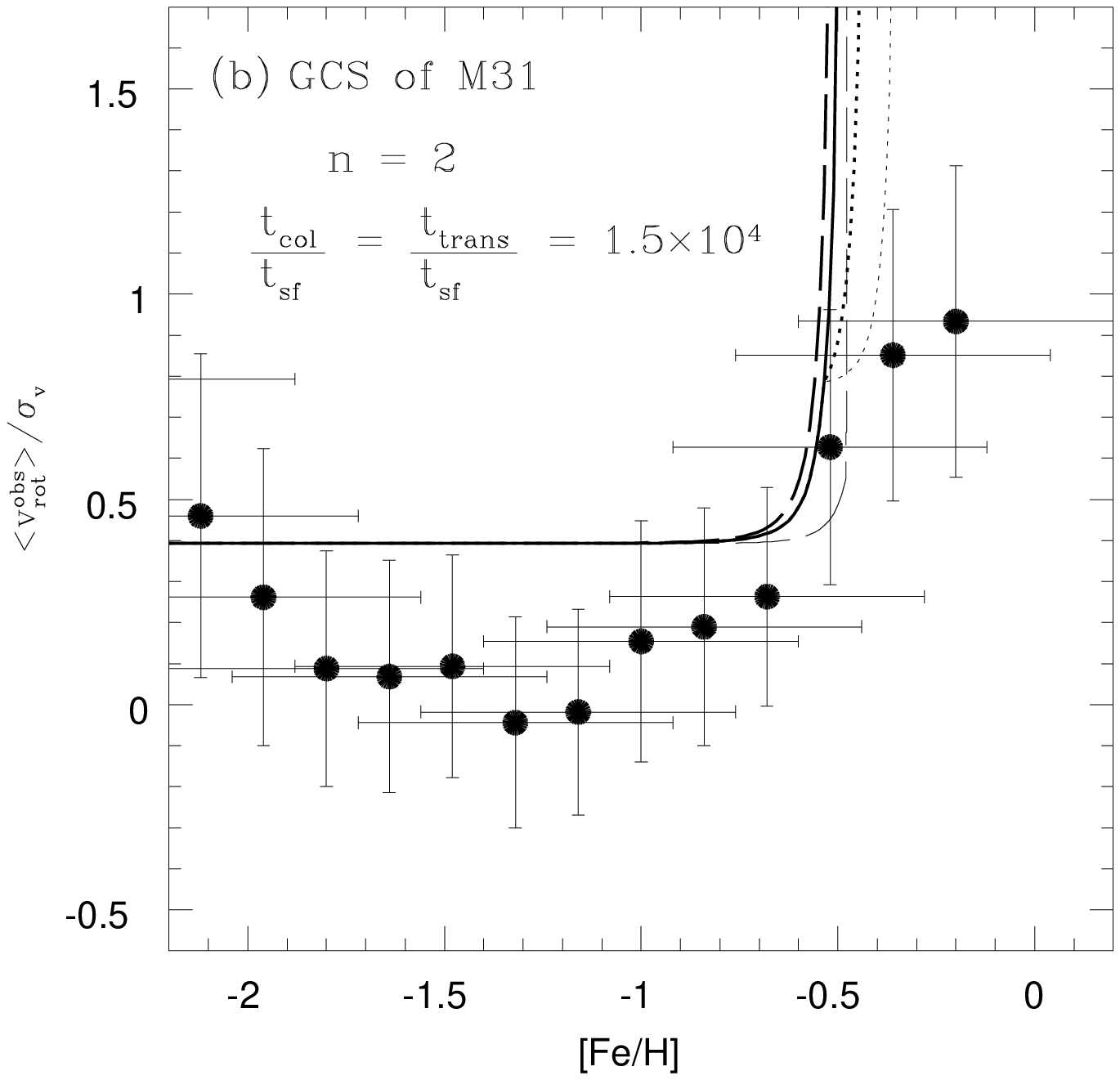]{
(a)Relation between the metallicity and normalized specific rotation of the GCS of M31.
The observational data are from Barmby et al. (2000)(solid circles).
The numerical model results are as described for Fig. 2(a), but for $t_{col}/t_{sf} = t_{trans}/t_{sf} = 15.0$.
(b)Similar diagram, but for n=2 and $t_{col}/t_{sf} = t_{trans}/t_{sf} = 1.5\times 10^{4}$.
\label{fig5}} 
\figcaption[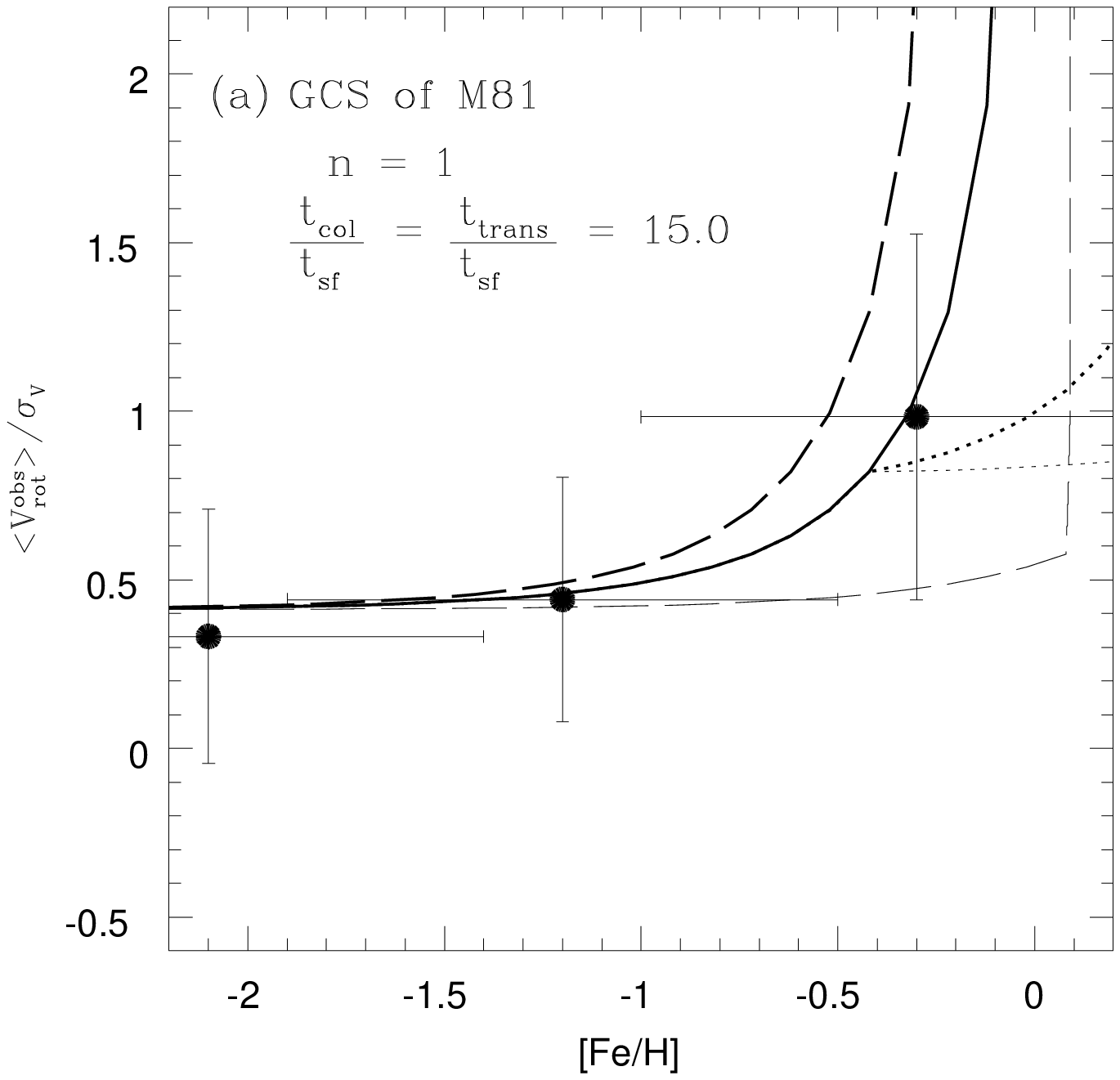, 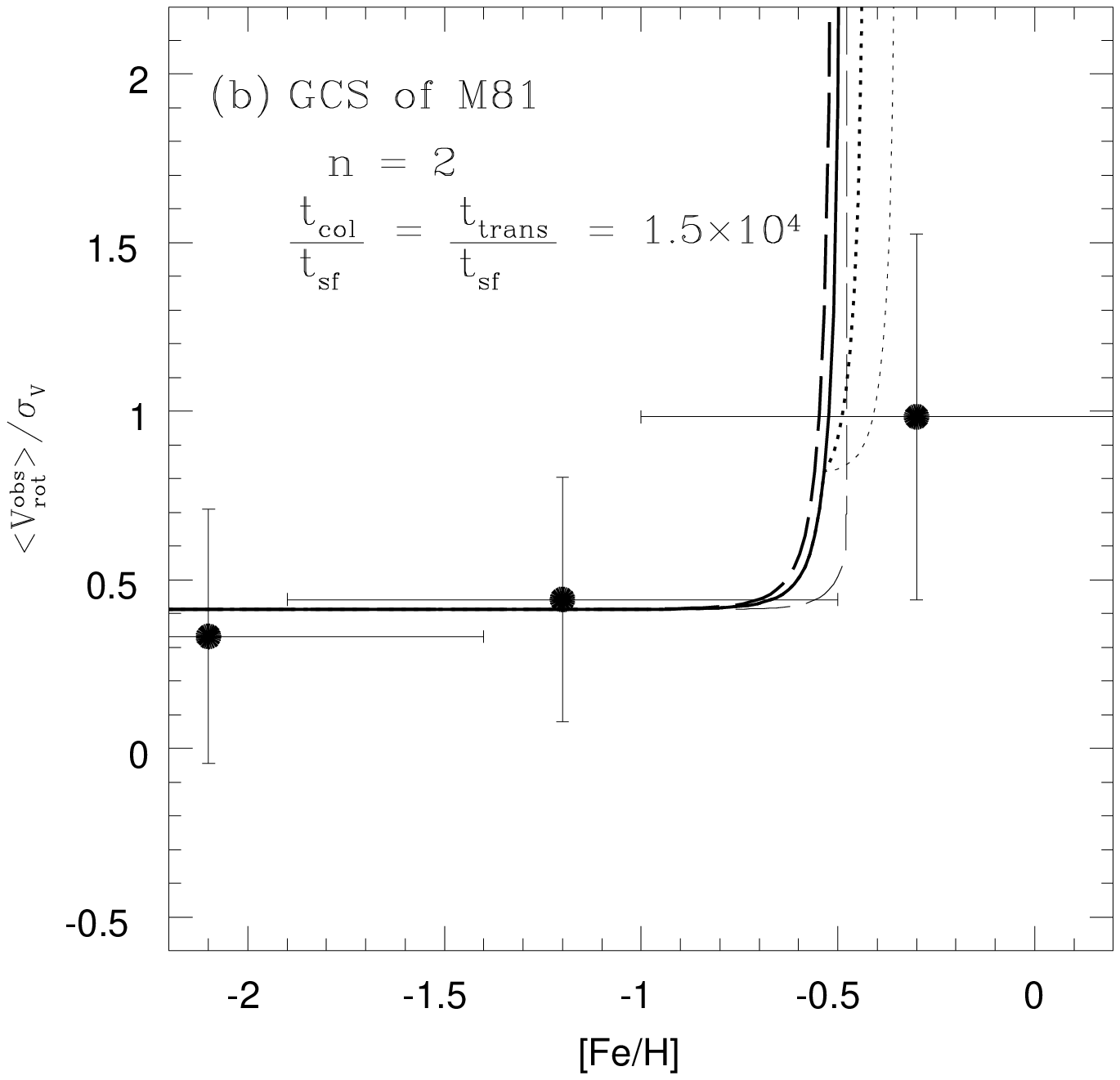]{
(a)Relation between the metallicity and normalized specific rotation of the GCS of M81.
The observational data are from Perelmuter, Brodie \& Huchra (1995)(solid circles).
The numerical model results are as described for Fig. 2(a), but for $t_{col}/t_{sf} = t_{trans}/t_{sf} = 15.0$.
(b)Similar diagram, but for n=2 and $t_{col}/t_{sf} = t_{trans}/t_{sf} = 1.5\times 10^{4}$.
\label{fig6}} 
\figcaption[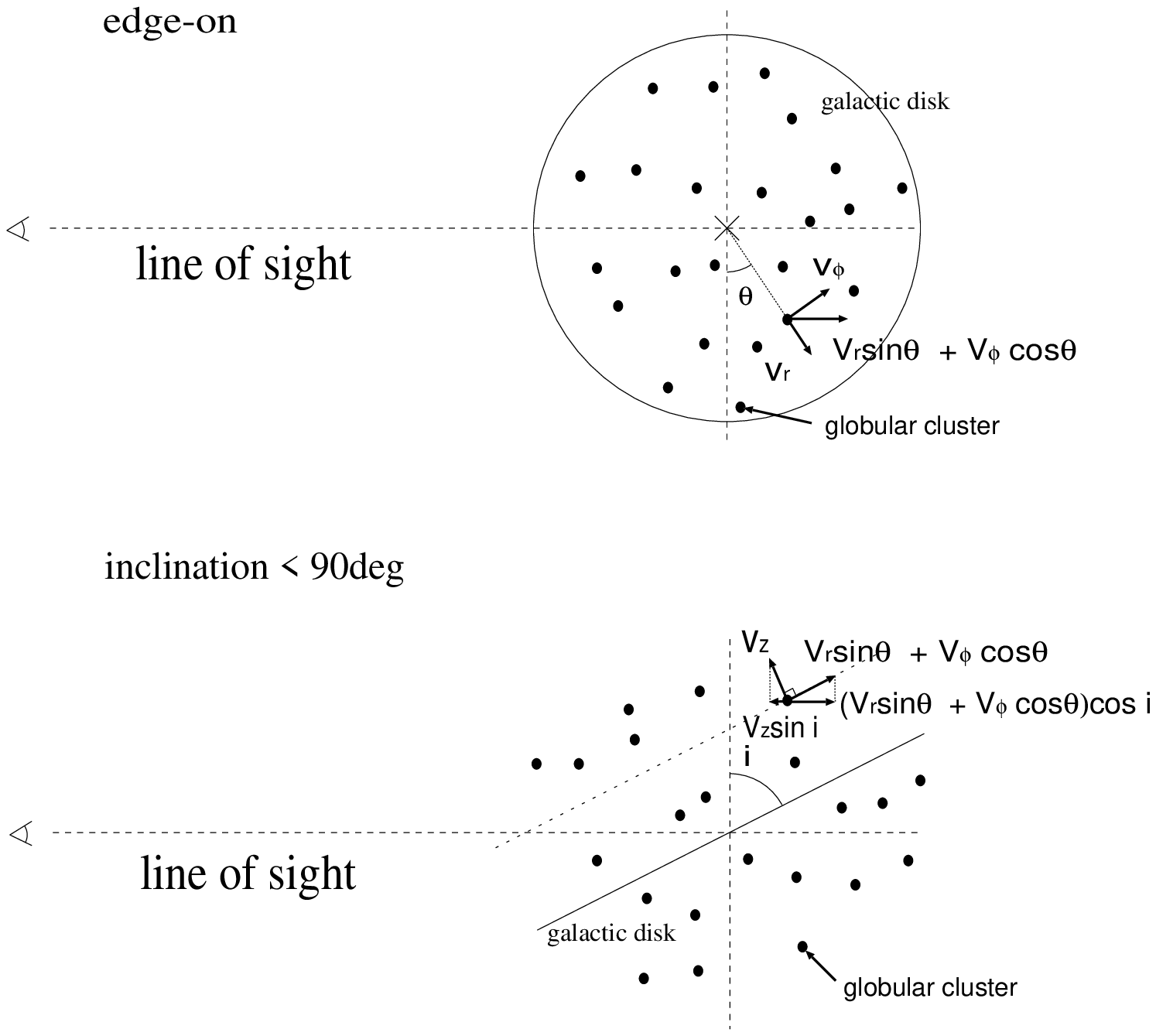]{
Observed velocity components of GCs of a disk galaxy.
The upper figure show the disk galaxy seen edge-on. The lower figure show the disk galaxy having inclination $< 90^{\circ }$.   
\label{fig7}} 
\figcaption[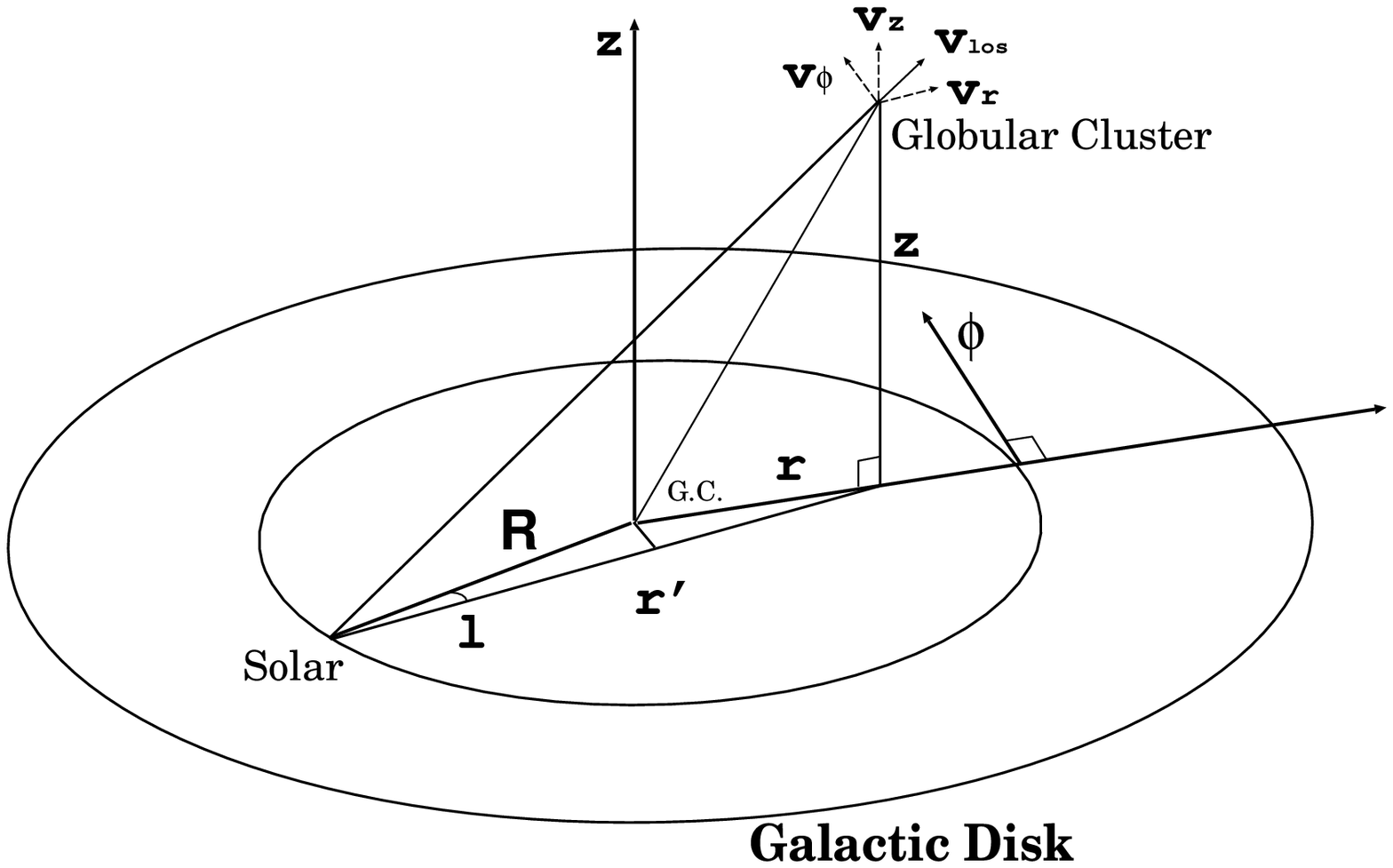]{
Observed velocity components of GCs of our Galaxy.
GCs are observed from the solar position $R = 8$kpc of Galactic disk.
\label{fig8}}

\clearpage

\begin{figure}
\plotone{f1.eps}
\end{figure}

\begin{figure}
\plottwo{f2a.eps}{f2b.eps}
\end{figure}

\clearpage

\begin{figure}
\plottwo{f3a.eps}{f3b.eps}
\end{figure}

\clearpage

\begin{figure}
\plottwo{f4a.eps}{f4b.eps}
\end{figure}

\clearpage

\begin{figure}
\plottwo{f5a.eps}{f5b.eps}
\end{figure}

\clearpage

\begin{figure}
\plottwo{f6a.eps}{f6b.eps}
\end{figure}

\begin{figure}
\plotone{f7.eps}
\end{figure}

\begin{figure}
\plotone{f8.eps}
\end{figure}

\end{document}